# Spatial Distribution of Metal Emissions in Supernova Remnant 3C 397 Viewed with *Chandra* and *XMM*

**Bing JIANG and Yang CHEN**

Department of Astronomy, Nanjing University, Nanjing, 210093, China

**Abstract**
We present X-ray equivalent width imaging of 3C 397 for Mg Heα, Si Heα, S Heα, and Fe Kα complex lines with *Chandra* and *XMM*-Newton observations. The images reveal that the heavier the element is, the smaller the extent of the element distribution is. The Mg emission is evidently enhanced in the southeastern blow-out region, well along the radio boundary there, and appears to partially envelope the eastern Fe knot. Two bilateral hat-like Si line-emitting structures are along the northern and southern borders, roughly symmetric with respect to the southeast-northwest elongation axis. An S line-emitting shell is located just inside the northern radio and IR shell, indicating a layer of reversely shocked sulphur in the ejecta. A few enhanced Fe features are basically aligned along the diagonal of the rectangular shape of the SNR, which implicates an early asymmetric SN explosion.

## 1. Introduction

Supernovae (SNe) are key factories to feed the universe with energy and heavy elements. Unfortunately, the physical mechanisms of nucleosynthesis product ejection in an SN event are still mysterious. Thanks to high-resolution space observations, the spatially resolved distribution of the ejecta provides some records of the imprints, allowing us to study stellar nucleosynthesis. For instance, relevant works on the youngest and brightest known Galactic supernova remnant (SNR) Cas A revealed a jet structure and the Fe-rich ejecta lying outside the Si-enriched material [1-4]. It was proposed that Cas A has undergone an asymmetric explosion and an inversion of the explosive O- and Si-burning products due to neutrino-driven convection during the initial phase of the SN explosion [1,4].

3C 397 (G41.1-0.3) is a radio and X-ray bright Galactic SNR with a peculiar rectangular morphology. The *ROSAT* HRI imaging displayed a bipolar X-ray shape with a bright central core [5-7]. Using the *ASCA* and *Chandra* observations, neither pulsed signal nor hard non-thermal emission was detected from the central spot, as would be expected from a pulsar and its wind nebula [5,7,8]. The *ASCA* observation found that the hot gas interior to the remnant is metal enriched [5,7]. Safi-Harb et al. [8] have performed a detailed spatially resolved spectral analysis with the *Chandra* data and found that the metal enhancement is almost everywhere within the SNR. A jet-like structure was also detected along the symmetry axis of the SNR [8].

Aiming to investigate the spatial distribution of the X-ray emission from metal species in SNR 3C 397, we revisited the *Chandra* data and analyzed the *XMM*–Newton data, and, in this note, present the Equivalent Width (EW) maps for bright metals in the remnant.

## 2. Data Analysis

SNR 3C 397 was observed by *Chandra* on 2001 September 6 (ObsID: 1042, PI: S. S. Holt), with the back-illuminated chip S3 of the Advanced CCD Imaging Spectrometer for an exposure of 66.0 ks. We reprocessed the data from level 1 similar to that in Safi-Harb et al. [8] except using the CIAO v3.01 software packages. The resulting effective exposure is 65.4 ks and the angular resolution is ~0.5".

Three pointings by *XMM*-Newton EPIC were taken on 2003 October 15, 2004 April 23 and 2004 April 25 (Obs. ID: 0085200301, 0085200401, and 0085200501, PI: Torii, Ken'ichi) with the thin filter and the full frame mode. The total exposure was 60.5 ks. Due to significant contamination on the EPIC-PN camera by cosmic rays, we only focused on the EPIC-MOS data for subsequent analyses. The EPIC-MOS data were calibrated using the software SAS v8.0.1, together with the latest calibration files. The time intervals with heavy proton flarings were excluded accordingly when the count rates exceeded about 20% above the average level. The remaining effective exposures are 32.6 ks, for both MOS1 and MOS2, and the angular resolutions are ~4".

Four metal complex lines of Mg Heα, Si Heα, S Heα,

†Send offprint requests to: Bing Jiang, bjiang@nju.edu.cn

and Fe Kα are distinct in the spectrum of the entire SNR (Figure 1). The EW images for these four lines were produced (Figure 2) after combining the processed *Chandra* and *XMM*-Newton data so as to improve the signal-to-noise ratio. We followed the methods described by Hwang et al.[2] and Park et al. [9]. For each atomic emission line, we extracted the images in the line bandpasses and in the two "shoulder" bands of the line bandpasses (see Table 1) from *Chandra* and *XMM* data, respectively. They were re-binned to a resolution of 4" with background subtracted and then the *Chandra* and *XMM* data were summed together in each corresponding band. The background in each band was averaged from local regions near the remnant. Assuming that the line-to-continuum ratio is invariable among all the observations, such a band-by-band summation is feasible according to the componendo. The widths of line bands were selected no less than the energy resolution of *Chandra* and *XMM* data. The continua were supposed to be linear around the line bands and were interpolated between the two shoulders for each pixel. The continuum-subtracted line intensity was then divided by the estimated continuum to generate the EW image. The adaptive mesh refinement was applied to achieve the minimum counts of 20 per pixel for the Mg, Si and S lines, and 10 counts per pixel for the Fe line. In order to remove the noise due to the background, we set the EW values to be zero where the line-to-continuum ratios were negative and where the estimated continuum fluxes were low. For comparison, the *Chandra* counts map of the 4-6 keV continuum is shown in Figure 3 to illustrate the line-free emission. Note that the determination of the underlying continua in the line profiles, for instance, Mg Heα, is somewhat artificial due to the strong extinction in the soft X-rays (≤1 keV) and the possible mixture with the emission of the Fe L lines. In Figure 4, we smoothed the EW images using a Gaussian function with a radius of 16".

## 3. Results

The EW maps (Figures 2 & 4) for Mg, Si, S and Fe show the spatial distribution of line emissions from these elements. They are different from each other and are also different from the line-free image (Figure 3). From Mg to Fe, the heavier the element is, the smaller the extent of the distribution is, roughly consistent with the nucleosynthesis onion model. The distribution of the elemental emission is described in detail below.

(1) **Mg** Magnesium is ubiquitous across the remnant. It is evidently enhanced in the southeastern blow-out region, well along the radio boundary there. There are some enhanced Mg line-emitting knots along the western radio shell. A bright arc structure of Mg is located to the north of the geometric center of the remnant.

(2) **Si** Bilateral hat-like structures of silicon are remarkable along the north and south borders which are weak in the X-ray continuum (Figure 3). The two structures seem to be roughly symmetric with respect to the SNR's southeast-northwest elongation axis.

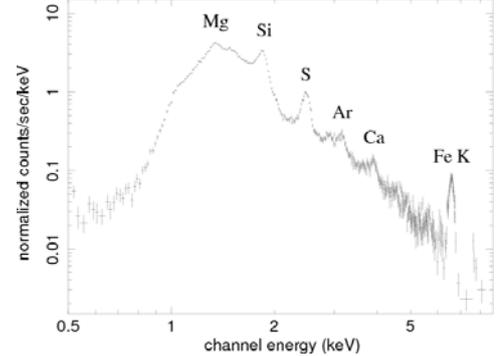

**Figure 1**. *Chandra* spectrum of the entire remnant, extracted from the region marked in Figure 3. The spectrum was grouped to achieve a background-subtracted signal-to-noise ratio of 3. The background was extracted locally from source-free regions around 3C 397.

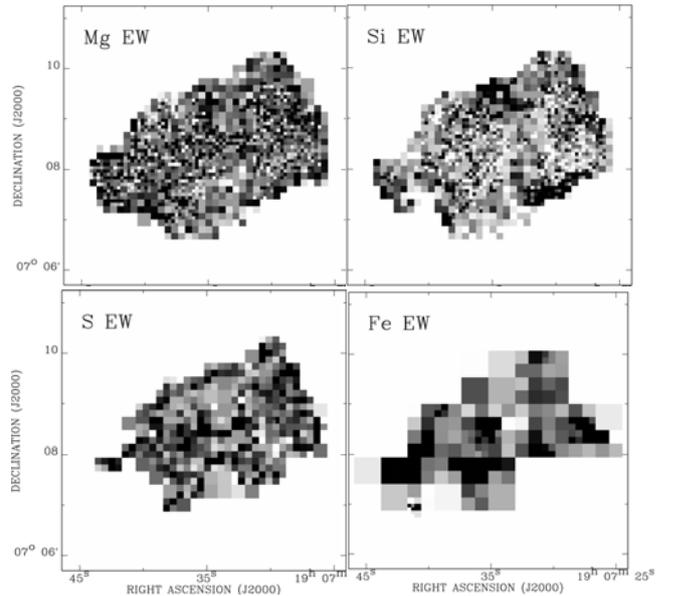

**Figure 2**. EW images for Mg Heα, Si Heα, S Heα, and Fe Kα lines (see text in Sec. 2).

**Table 1** Energy bands used for the EW images

| Species | Left shoulder (keV) | Line (keV) | Right shoulder (keV) |
|---|---|---|---|
| Mg | 1.08-1.22 | 1.26-1.4 | 1.55-1.69 |
| Si | 1.55-1.69 | 1.75-1.93 | 2.00-2.10 |
| S | 2.00-2.10 | 2.35-2.55 | 2.60-2.80 |
| Fe Kα | 6.00-6.25 | 6.35-6.85 | 6.95-7.20 |

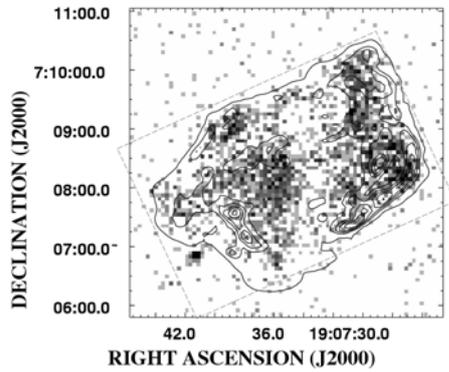

**Figure 3.** *Chandra* counts map for 4-6 keV continuum, overlaid by VLA 1.4 GHz radio continuum emission in contours with levels of 2, 9, 13.5, 18, 23, 32 mJy beam$^{-1}$. The grey dashed box denotes the region from where the spectrum in Figure 1 was extracted.

(3) **S** An S line-emitting shell is seen along, but *just inside of*, the northern border that is indicated by the radio and IR shell (see Jiang et al. 2009[a]). Jiang et al. (2009)[a] discovered a molecular density jump along the northern border. Therefore, this S shell may indicate the reversely shocked sulphur in the ejecta. Two S-line emitting arcs are located near the geometric center; note that the Mg emission happens to be weak along the southern S arc. In addition, the S emission is weak basically along the bright Mg arc to the north of the center. Furthermore, the distribution of S appears to be different from that of Si, not showing a close correlation seen in the O-rich SNR Cas A between the two elements[10].

(4) **Fe** There are two Fe line-emitting knots near the center; the northern one is partially coincident with the "hot spot" in Safi-Harb et al.[8]. Another bright Fe line-emitting knot is seen near the eastern boundary, seemingly enveloped in part by the eastern Mg line-emitting shell. Fe emission is also enhanced near the western rectangular corner, where the emissions of Mg and Si both happen to be weak. This location seems to be coincident with the westmost $^{12}$CO (J=1-0) line broadened region (Jiang et al. 2009[a]), implying that the molecular gas there might be impacted by the Fe-rich ejecta. These bright Fe features are basically aligned along the diagonal of the rectangular shape of the SNR, which implicates an early asymmetric SN explosion.

The bright point source to the southeast and outside of the remnant in the Fe EW map and in Figure 3 is highly unlikely to be associated with 3C 397 and was suggested to most likely be a nearby AGN[8]. Notably, it disappears in the *XMM* data. Such a significant magnitude of the flux variation (a factor of ~300) between the two observations (over about two years) seems to conflict with the AGN origin scenario, but is likely to be a transient high-mass X-ray binary.

*Note added in proof* -- We note that an imaging and spectroscopic study of the XMM-Newton data is also conducted independently by another group (Safi-Harb et al. 2009, in preparation for ApJ; private communication).

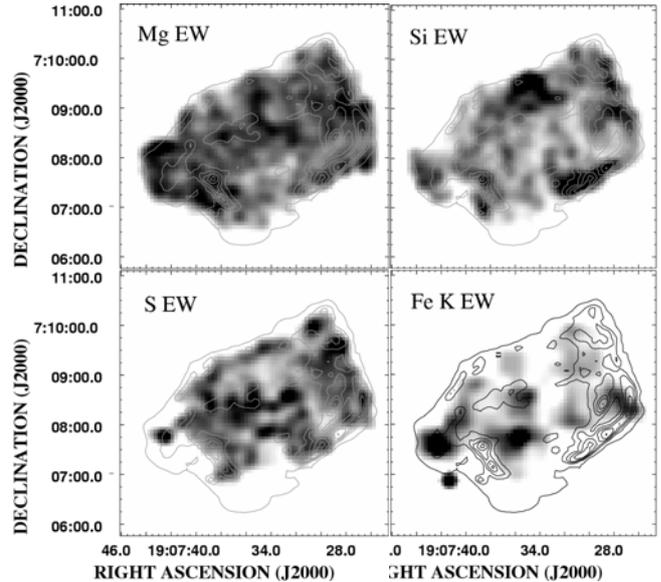

**Figure 4.** Smoothed EW images of Figure 2 with a resolution of 16". The contours are the same with that in Figure 3.

**Acknowledgments** We thank Tracey Delaney for helpful discussion and Lawrence Rudnick for providing the VLA data of SNR 3C397. Y.C. acknowledges support from NSFC grants 10725312 and10673003 and the 973 Program grant 2009CB824800.

---

[a] Cavity of molecular gas associated with supernova remnant 3C 397, Jiang B, Chen Y, Wang J Z, Su Y, Zhou X, Safi-Harb S, Delaney T, 2009, submitted.